\documentclass{appolb}
\usepackage{graphicx}
\newcommand{\inst}[1]{$^{#1}$}

\newcommand{\apj}{ApJ}
\newcommand{\mnras}{MNRAS}

\newcommand{\aap}{A{\&}A}

\newcommand{\apjs}{ApJS}

\begin{document}
\title{Magnetorotational Instability in
  Core-Collapse Supernovae\thanks{Talk presented by T.~Rembiasz at the 3rd Conference of the Polish Society on Relativity, Krak\'ow, Poland, September 25–-29, 2016.}
}
\author{T.\,Rembiasz\inst{1},
  M.\,Obergaulinger\inst{1},
  J.\,Guilet\inst{2,3},
  P.\,Cerd\'a-Dur\'an\inst{1}, 
  M.A.\,Aloy\inst{1},
  E.\,M{\"u}ller\inst{2}
  \\
\address{
\inst{1} Departamento de Astronom\'{\i}a y Astrof\'{\i}sica,  Universidad de Valencia, \\  C/ Dr.~Moliner 50, 46100 Burjassot, Spain  \\
\inst{2} Max-Planck-Institut f{\"u}r Astrophysik, Karl-Schwarzschild-Str.~1, \\ 85748 Garching, Germany \\
\inst{3}  Max Planck/Princeton Center for Plasma Physics
} 
}
\maketitle
\begin{abstract}
  We discuss the relevance of the magnetorotational instability (MRI)
  in core-collapse supernovae (CCSNe). Our recent numerical studies
  show that in CCSNe, the MRI is terminated by parasitic instabilities
  of the Kelvin–Helmholtz type. To determine whether the MRI can
  amplify initially weak magnetic fields to dynamically relevant
  strengths in CCSNe, we performed three-dimensional simulations of a
  region close to the surface of a differentially rotating
  proto-neutron star in non-ideal magnetohydrodynamics with two
  different numerical codes. We find that under the conditions
  prevailing in proto-neutron stars, the MRI can amplify the magnetic
  field by (only) one order of magnitude. This severely limits the
  role of MRI channel modes as an agent amplifying the magnetic field
  in proto-neutron stars starting from small seed fields. 
\end{abstract}
%\PACS{PACS numbers come here}
  
\section{Introduction}

The  magnetorotational instability (MRI)  was suggested 
by  \cite{Balbus_Hawley__1991__ApJ__MRI} 
to be the physical mechanism driving the redistribution of angular momentum required for the
accretion process in Keplerian discs orbiting compact objects. 
Reference \cite{Akiyama_etal__2003__ApJ__MRI_SN} pointed out the potential
importance of the MRI for rapidly-rotating core collapse supernovae
(CCSNe) where it may amplify the weak pre-collapse fields to a
  dynamically relevant strength, thereby generating
  magnetohydrodynamic (MHD) turbulence and making rotational energy
  available for launching an explosion. Their simplified simulations
as well as multi-dimensional models (e.g.\,
\cite{Obergaulinger__2006,Cerda__2008}) showed that in such CCSNe,
proto-neutron stars (PNSs) possess regions where the MRI can grow on
shorter time scales than the time between the bounce and the
successful explosion.  However, due to limitations of the above
mentioned numerical simulations, the question whether the MRI could
amplify the initial magnetic field to dynamically important field
strengths so that it could tap the rotational energy of the core and
power MHD turbulence remained open.

An upper limit for the magnetic field amplification caused by the MRI
can be given by assuming that the MRI ceases to grow once the magnetic
field comes close to equipartition with the energy of the differential
rotation.  In a CCSN, this would correspond to dynamically important
field strengths up to $10^{15} \, \mathrm{G}$.  However,
\cite{Goodman_Xu} studied the phase of exponential
  growth of the MRI which is characterised by channel modes,
  i.e.\ layers of radial and azimuthal  magnetic field and velocity with alternating
  polarity and found that they can be 
 unstable against
\emph{secondary} (or \emph{parasitic}) \emph{instabilities} of
Kelvin-Helmholtz (KH) or tearing-mode (TM) type. Hence according to
the model of parasitic instabilities (further developed and studied
analytically by \cite{Latter_et_al,Pessah}), the MRI channel modes
could be disrupted by secondary instabilities before the equipartition
of the energy of the magnetic fields and of the differential rotation
is reached.

The goal of numerical studies of  \cite{Rembiasz_et_al_2016_mri_i,Rembiasz_et_al_2016_mri_ii,Rembiasz_et_al_2016_mri_astr}
was to answer whether the MRI channel modes can
amplify the magnetic field to relavant strengths  in CCSNe as well as  to test the theoretical predictions of the parasitic model.
We summarise the findings of those papers in the next section.

\section{Numerical simulations}

In \cite{Rembiasz_et_al_2016_mri_i}, following
  \cite{Obergaulinger_et_al_2009}, we performed 
two-dimensional (2D)  and  three-dimensional (3D) 
  shearing-disc (semi-global) simulations which focus on a small representative region
  of a PNS. The simulations were done with  the finite volume code \textsc{Aenus}
\cite{Obergaulinger__2008__PhD__RMHD} in  resistive-viscous  magnetohydrodynamics.
 According to the estimates of   \cite{Guilet_et_al_2015}, close to the surface of PNS, the
  contribution of neutrinos to viscosity is low
 and therefore the flow  is characterised there by  high   Reynolds numbers.
Therefore, in our studies \cite{Rembiasz_et_al_2016_mri_i},  we mainly focused
on the limit of  high   Reynolds numbers, which required using  very fine grids.
 We point   out that it is very important to distinguish the effects of
  numerical viscosity and resistivity from their physical counterparts.
  Therefore, we studied numerical errors giving rise to artificial
  dissipation in detail \cite{Rembiasz_et_al_2017}, finding that the
  use of numerical methods of very high convergence order is crucial
  for resolving the small-scale features of the parasitic
  instabilities \cite{Rembiasz_et_al_2016_mri_astr}. 

\begin{figure}
\centerline{%
\includegraphics[width=0.49\textwidth]{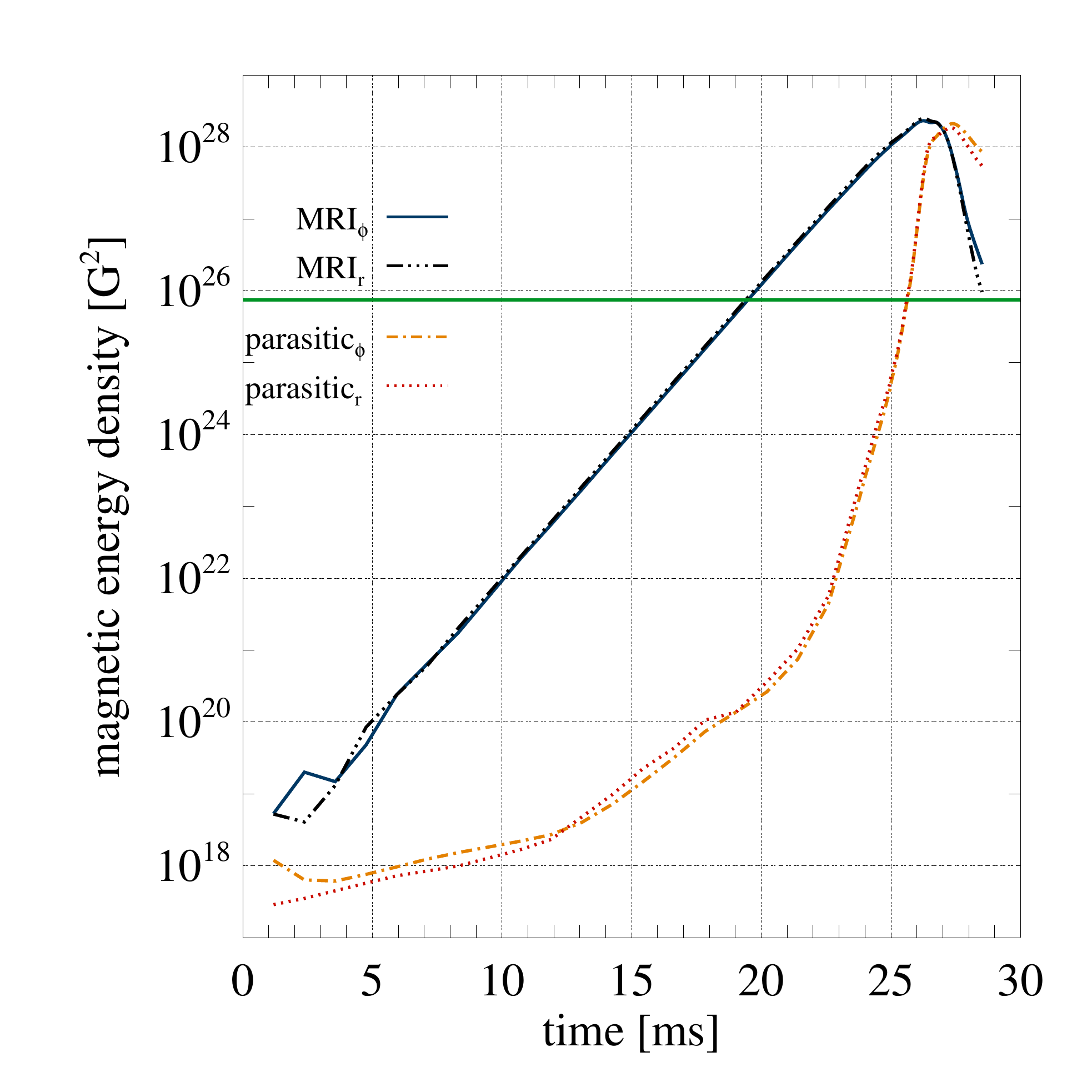}
\includegraphics[width=0.49\textwidth]{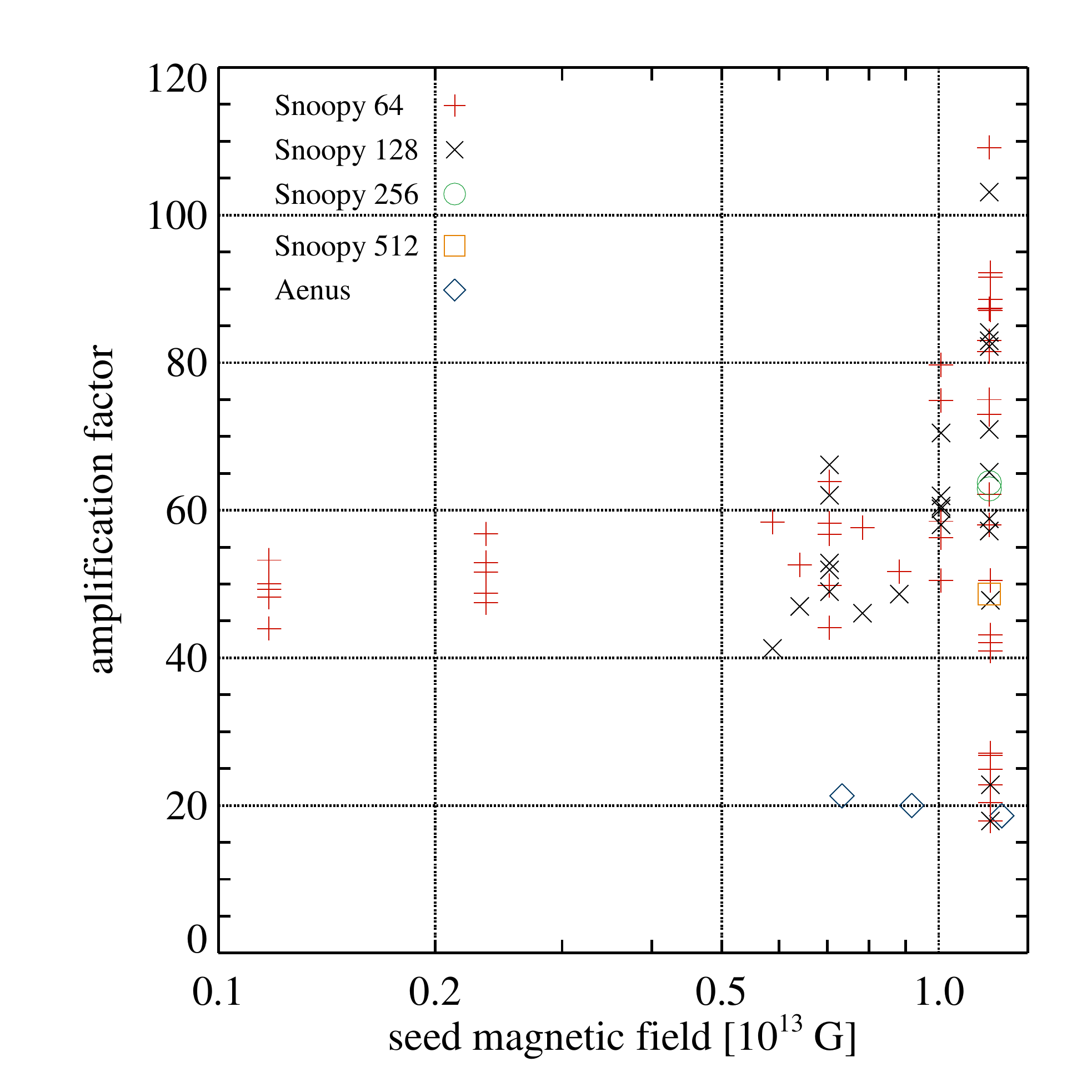}}
\caption{\emph{Left}:  magnetic field energy of  the MRI channel
  modes (black and blue) and of  the parasitic instabilities (red and
  orange) as a function of time in a simulation done with
  \textsc{Aenus}, with an initial (seed) magnetic field strength
 $1.22  \times 10^{13}\,\mathrm{G}$ (i.e.\, magnetic field energy
 density $7.44  \times 10^{25}\,\mathrm{G}^2$; marked with a green
horizontal line). When the energy  of both instabilities is
comparable, the  MRI is terminated. 
\emph{Right}:   magnetic field amplification 
(roughly defined as the ratio of the amplitudes of the initial seed magnetic
field to the amplitude of the MRI channel at termination)
as a function of initial magnetic field strength in simulations performed with \textsc{Aenus}
and \textsc{Snoopy}. 
See  \cite{Rembiasz_et_al_2016_mri_i} for more details.
   }
\label{fig:fig01}
\end{figure}

Our main result is that in 3D simulations with high Reynolds numbers, 
  the MRI growth is, in accordance with the parasitic model,  terminated
by secondary parasitic KH instabilities, whose properties are in a
good agreement with the theoretical predictions.
2D simulations, because of the axisymmetry constraint, 
give a qualitatively wrong result, i.e.\,
the MRI is terminated by TMs (as already observed by  \cite{Obergaulinger_et_al_2009}),

In \cite{Rembiasz_et_al_2016_mri_ii}, we used two numerical codes,
i.e.\  \textsc{Aenus}
 and the pseudo spectral code
\textsc{Snoopy} (\cite{lesur05}; using the shearing box and
incompressible approximations) to test the prediction of the parasitic
model for the maximum amplification of the magnetic field by the MRI
channel modes (see Fig.\, \ref{fig:fig01}).  We found a disagreement
between the theoretical predictions and scaling laws for the
  termination field strength obtained from our simulations.  With the
help of our scaling laws, we estimated that under the conditions found
in PNSs a realistic value for the magnetic field amplification is of
the order of $10$.

This casts doubt on the viability of the MRI channel modes as
an agent amplifying the magnetic field in proto-neutron stars starting
from small seed fields. A further amplification should therefore rely
on other physical processes, such as for example an MRI-driven
turbulent dynamo (for numerical studies in the presence of buoyancy,
see \cite{Guilet_Mueller_2015}) or the standing accretion shock
instability.

\section*{Acknowledgments}

M.A, P.C.D., M.O. and T.R.\  acknowledge support from
the European Research Council (grant CAMAP-259276) as well as from grants 
AYA2013-40979-P, AYA2015-66899-C2-1-P and
PROMETEOII/2014-069.  J.G.\  and  E.M.\  acknowledge support from the Max-Planck-Princeton Center
for Plasma Physics. The computations were performed at the Leibniz
Supercom- puting Center of the Bavarian Academy of Sciences and
Humanities (LRZ), the Max Planck Computing and Data Facility (MPCDF)
and at the Servei d'Inform\'atica of the University of Valencia.

\end{document}